\documentclass[a4paper,10pt]{article}
\usepackage[utf8]{inputenc}
\usepackage[english]{babel}
\usepackage{cite}
\usepackage{graphicx}
\usepackage{csquotes}
\usepackage{subcaption} 
\usepackage{rotating} 
\usepackage{gensymb}
\usepackage{array}
\usepackage{float}
\usepackage{esint}
\usepackage{url}
\usepackage{epstopdf}
\usepackage{epsfig}
\usepackage{placeins}
\usepackage{amssymb}
\usepackage{braket}
\usepackage{threeparttable} 
\usepackage{multirow}
\usepackage{booktabs}

\usepackage[top=0.7in, bottom=0.7in, left=0.5in, right=0.5in]{geometry}

\usepackage{setspace}
\makeatletter

\newcommand{\Rmnum}[1]{\expandafter\@slowromancap\romannumeral #1@}
\makeatother

\newcommand{\ra}[1]{\renewcommand{\arraystretch}{#1}}

\usepackage{authblk}

\title{\Large Stellar 30-keV neutron capture in $^{94,96}$Zr and the $^{90}$Zr$(\gamma,n)^{89}$Zr photonuclear reaction with a high-power liquid-lithium target}

\date{}

  \author[a]{M. Tessler}
  \author[a,\footnote{Corresponding author: Michael Paul, Racah Institute of Physics, Hebrew University,
Jerusalem, Israel 91904, T: +972-2-6584795, F: +972-2-6586347
\newline Email address: paul@vms.huji.ac.il}] {M. Paul}
  \author[a,b]{A. Arenshtam}
  \author[a,b]{G. Feinberg}
  \author[a]{M. Friedman}
  \author[a,b]{S. Halfon}
  \author[b]{D. Kijel}
  \author[b]{L. Weissman}
  \author[b]{O. Aviv}
  \author[b]{D. Berkovits}
  \author[b]{Y. Eisen}
  \author[b]{I. Eliyahu}
  \author[b]{G. Haquin}
  \author[b]{A. Kreisel}
  \author[b]{I. Mardor}
  \author[b]{G. Shimel}
  \author[b]{A. Shor}
  \author[b]{I. Silverman}
  \author[b]{Z. Yungrais}
 \affil[a]{\textit{Racah Institute of Physics, Hebrew University, Jerusalem, Israel 91904}}
 \affil[b]{\textit{Soreq NRC, Yavne, Israel 81800}}

\begin{document}
\clearpage\maketitle
\thispagestyle{empty}

\section*{Abstract}

\noindent A high-power Liquid-Lithium Target (LiLiT) was used for the first time for neutron production via the thick-target $^7$Li$(p,n)^7$Be reaction and quantitative determination
of neutron capture cross sections. Bombarded with a 1-2 mA proton beam at 1.92 MeV from the Soreq Applied Research Accelerator Facility (SARAF),
the setup yields a 30-keV quasi-Maxwellian neutron spectrum with an intensity of 3-5 $\times10^{10}$ n/s, more than one order of magnitude larger than present near-threshold $^7$Li$(p,n)$ neutron sources.
The setup was used here to determine the 30-keV Maxwellian averaged cross section (MACS) of $^{94}$Zr and $^{96}$Zr as $28.0\pm0.6$ mb and $12.4\pm0.5$ mb respectively,
based on activation measurements. The precision of the cross section determinations results both from the high neutron yield and from detailed simulations of the
entire experimental setup. We plan to extend our experimental studies to low-abundance and radioactive targets. In addition, we show  here that the setup yields intense high-energy
(17.6 and 14.6 MeV) prompt capture $\gamma$ rays from the $^7$Li$(p,\gamma)^8$Be reaction
with yields of $\sim 3\times10^8$ $\gamma$/s/mA and $\sim 4\times10^8$ $\gamma$/s/mA, respectively, evidenced by the $^{90}$Zr$(\gamma,n)^{89}$Zr photonuclear reaction.

\noindent \textit{Keywords:} $^7$Li$(p,n)$, high-intensity neutron source, Maxwellian Averaged Cross Section (MACS), $^{94,96}$Zr$(n,\gamma)$, $^{90}$Zr$(\gamma,n)$
\noindent\makebox[\linewidth]{\rule{\textwidth}{0.5pt}} 

The availability of high-intensity (mA range) linear accelerators \cite{mA,SARAF} and the recent commissioning of the high-power (kW range) Liquid-Lithium Target (LiLiT \cite{LiLiT1,LiLiT2})
pave the way to a new generation of experimental investigations in nuclear physics and astrophysics.
The $^7$Li$(p,n)^7$Be reaction just above the $^7$Li$(p,n)$ threshold (E$_p$ = 1.880 MeV) has been traditionally used to produce neutrons
in the epithermal energy regime; the thick-target angle-integrated neutron yield is known to have an energy distribution similar
to that of a flux of Maxwellian neutrons $v\cdot dn_{MB}/dE_n \propto E_n exp(-E_n/kT)$ at $kT\sim25$ keV \cite{Rat_Kep}.
It has been used in particular for the study of $s$-process nucleosynthesis (see \cite{s_proc} for a review)
but the neutron yield ($\lesssim 10^9$ n/s) was so far limited by the beam power dissipation in a solid Li (or Li-compound) target.
An increase in available neutron intensity and flux is considered an important goal \cite{Franz} in order to extend experimental investigations to low-abundance and radioactive targets.
We report here on first activation measurements of neutron capture cross sections by activation and their extrapolation to Maxwellian-Averaged Cross Sections (MACS)
in stable Zr isotopes using the high-intensity (1-2 mA) continuous-wave proton beam from the superconducting linear accelerator SARAF \cite{SARAF}
and the Liquid-Lithium Target (LiLiT \cite{LiLiT1,LiLiT2}). The neutron yield of $\sim$3-5$\times10^{10}$ n/s is
30-50 times larger than in existing facilities based on the near-threshold $^7$Li$(p,n)$ reaction for neutron production.
Preliminary results of these experiments were reported recently \cite{nic13,gitai_thesis}.
The choice of a $^{nat}$Zr target for these first experiments was motivated by the importance of the Zr isotopes along the path of $s$-process nucleosynthesis \cite{s_proc}.
Notably, Zr isotopic anomalies (relative to Solar abundances) detected in presolar grains are attributed to materials synthesized by the $s$-process at various stages of
He core and inter-shell burning in Asymptotic Giant Branch stars, contributing a complex patchwork of nucleosynthesis information \cite{Nicolussi,Zinner_98}.
A detailed analysis of these data in terms of astrophysical models \cite{Lugaro_03,Lugaro} emphasizes the importance of the $s$-process neutron capture cross sections
in this region of nuclides; several sets of experimental values of the neutron capture cross sections are available in the literature
\cite{Macklin,Boldeman,Musgrove,Wyrick,Tok90,Tagliente94,Allen,Tagliente96,Kadonis}. It was also shown recently \cite{s_temp} that the ratio of Zr to Nb abundances, $\frac{N(Zr)}{N(Nb)}$,
in $s$-process enriched stars (S-stars) can be used to estimate the relevant stellar temperatures.
We report here on new MACS ($kT$ = 30 keV) determinations
for the $^{94,96}$Zr isotopes. In our experiments, the liquid-lithium target bombarded by a high-intensity proton beam yields also intense
($\sim$ 7$\times10^8$ $\gamma$/s/mA) high-energy prompt gamma rays (17.6 and 14.6 MeV) from the $^7$Li$(p,\gamma)^8$Be capture reaction, evidenced by activation
through the $^{90}$Zr$(\gamma,n)^{89}$Zr photonuclear reaction.

\begin{figure}[h]
\centering
\includegraphics[width=0.5\columnwidth]{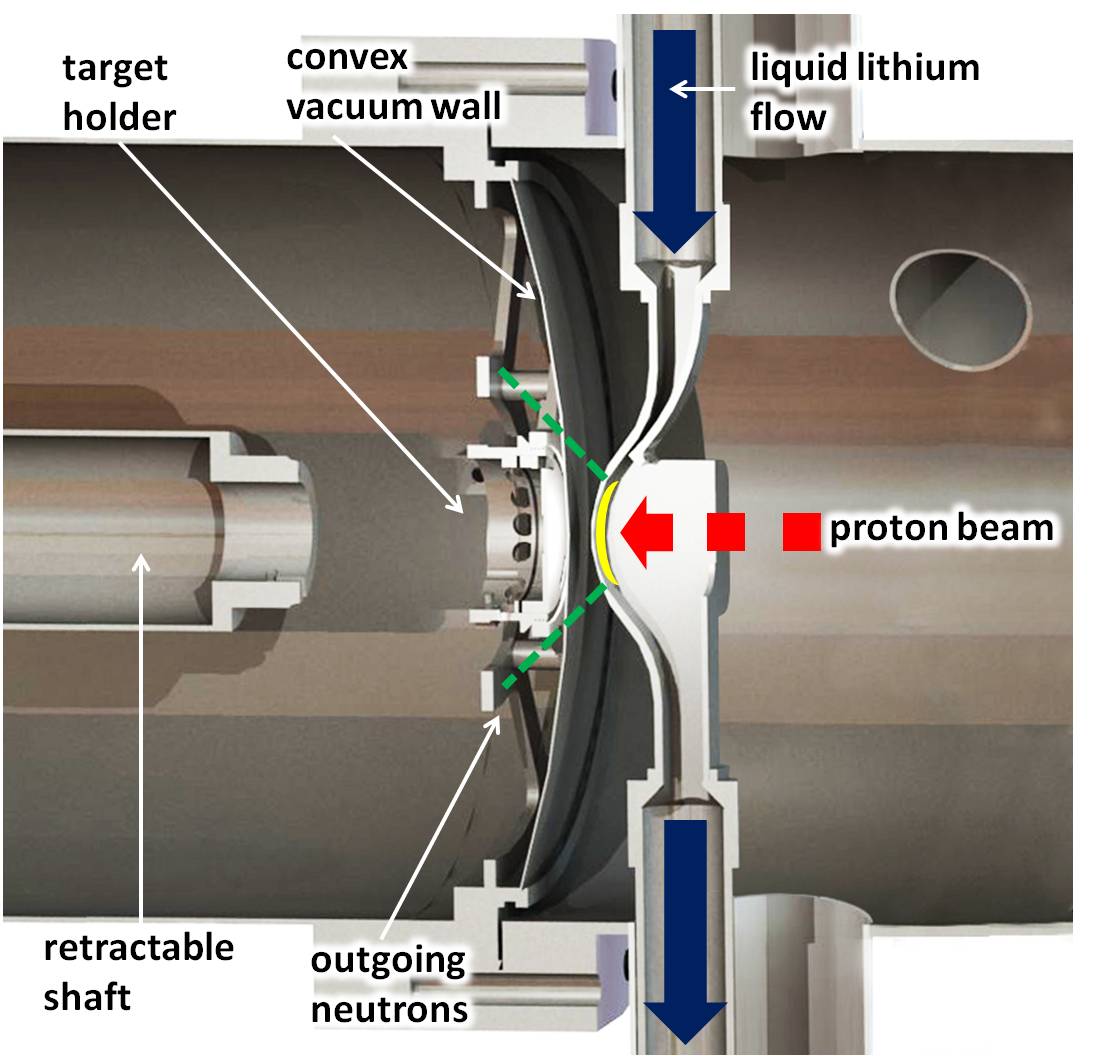}
  \caption{\label{fig:ex_setup} Detail diagram of the Liquid-Lithium Target (LiLiT) and activation target assembly.
  The (1-2 mA, $\sim$10 mm diameter) proton beam (dashed red arrow) impinges on the free-surface lithium film (yellow ellipse).
  The solid blue arrows show the inlet and outlet of the external circulating loop (see \cite{LiLiT1} for details).
  The activation target sandwich (Au-Zr-Au) is mounted on a circular holder and positioned in the outgoing neutron cone (green dotted lines)
  at a distance of 6-8 mm from the lithium surface in a vacuum chamber
  separated from the LiLiT chamber by a 0.5 mm stainless steel wall convex to the beam.
  The retractable shaft (at left) is used to load and unload rapidly the target assembly.}
\end{figure}

\begin{figure}[h]
\centering
 \includegraphics[width=0.5\columnwidth]{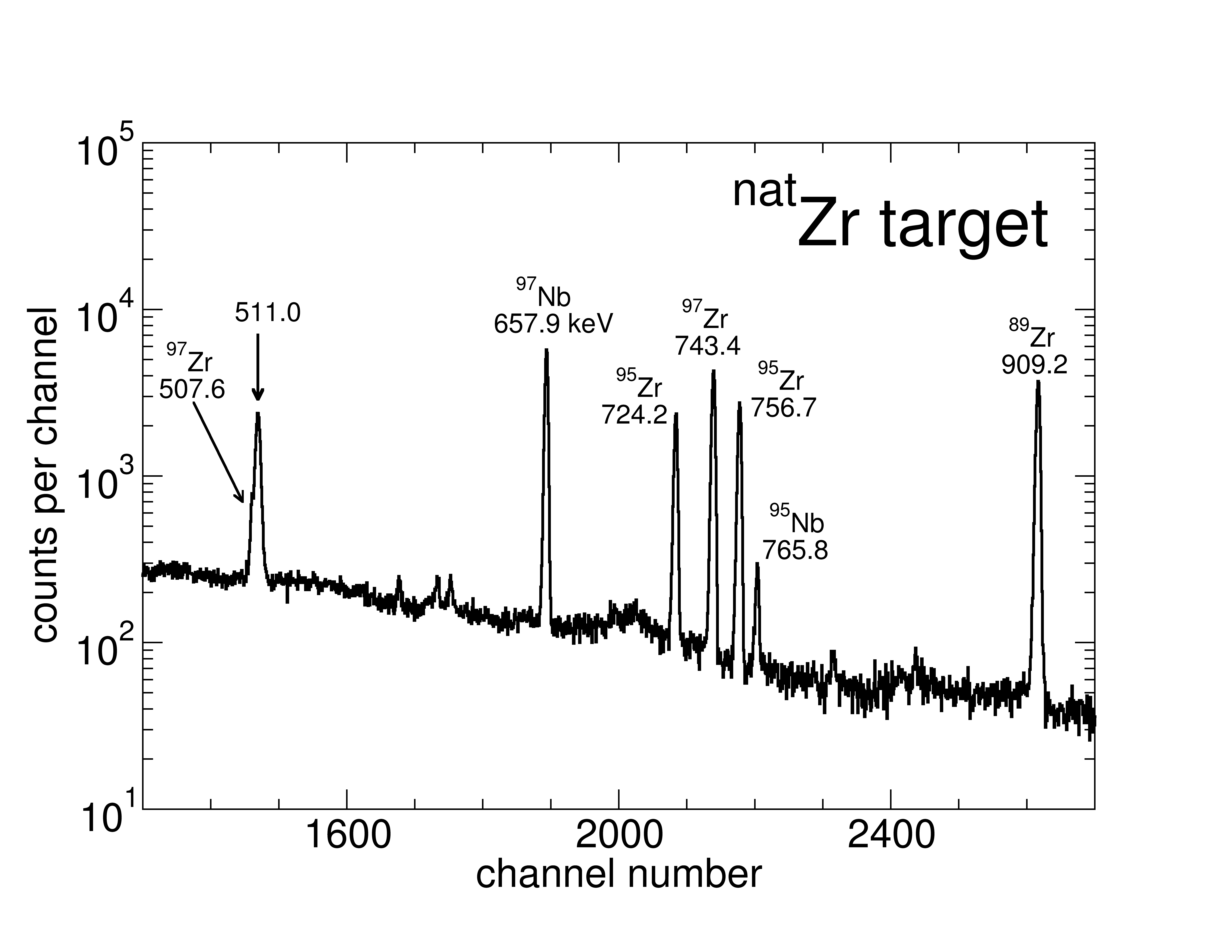}
 \caption{\label{fig:gamma_spec_old} $\gamma$-ray spectrum obtained by measuring the Zr sample for 226348 s, starting 83700 s after end of Exp \Rmnum{1}
 (see Table \ref{tab: MACS}) with a shielded HPGe. The photo-peaks from the decay of the activated Zr isotopes and daughters are labeled in keV.}
 \end{figure}

The Liquid-Lithium Target (LiLiT) consists of a film of liquid lithium ($\sim200\degree$C), 1.5 mm thick and 18 mm wide, forced-flown at high velocity ($\sim$ 2-5 m/s) onto a concave thin (0.3 mm)
stainless steel wall by an electromagnetic induction pump (fig. \ref{fig:ex_setup}; see \cite{LiLiT1} for details of the target design).
The (windowless) lithium film, bombarded by a $\sim$1-2 mA proton beam (E$_p \sim$ 1.92 MeV) from the SARAF accelerator \cite{SARAF} acts as both the neutron-producing target
and the power beam dump for the mA-proton beam ($\sim$ 2-3 kW) by fast transport of the liquid lithium to a reservoir and heat exchanger.
With the proton beam focused transversally to an approximate radial Gaussian distribution ($\sigma_r \sim$ 2.8 mm) in order to increase the neutron output flux,
the power volume density continuously deposited by the beam at the Bragg peak depth ($\sim$ 170 $\mu$m) in the liquid lithium is of the order of 1 MW/cm$^3$ \cite{LiLiT1}
while maintaining stable temperature and vacuum conditions with a lithium flow velocity of $\sim$2.5 m/s.
In the experiments described here, we activated a $^{nat}$Zr target positioned in a separate vacuum chamber behind a thin stainless wall (0.5 mm) of opposite curvature to that
of the liquid-lithium duct (fig. \ref{fig:ex_setup}), reducing thus the distance from the neutron source; the distance of the 25-mm diameter Zr target can be as small as
$6 \pm 1$ mm, intercepting a large fraction ($> 90\%$) of the outgoing neutrons.
Table \ref{tab: MACS} lists the activation conditions in three independent runs and their results. In each activation run,
the Zr target was tightly sandwiched by two Au foils of the same diameter serving
as neutron fluence monitors by off-line $\gamma$ counting of the $^{198}$Au activity. A $\gamma$ autoradiograph of the activated Au foils \cite{nic13} allowed us to determine
the centering of the neutron beam relative to the target assembly; correction for a slight misalignment ($\sim$2-3.5 mm),
observed in the different experiments and due to the difficulty
in precise steering of the high-intensity proton beam, was taken into account in the analysis. The proton beam energies, measured with concordant results
by Rutherford back scattering off a Au target after the acceleration module and by a time-of-flight pick-up system, were found to be slightly different in the three experiments
(see Table \ref{tab: MACS}) due to different tuning of the linear accelerator. In some of the experiments, the energy calibration was confirmed by a scan of the
narrow $^{13}$C$(p,\gamma)$ resonance ($E_p(lab)=1.746$ MeV) and the $^7$Li$(p,n)$ threshold region. A beam energy spread of $\sim$15 keV, estimated from beam dynamics calculations,
was verified experimentally \cite{gitai_thesis}. The activities of the Zr targets were measured (fig. \ref{fig:gamma_spec_old}) in the same geometry as the Au monitors
with a High-Purity Ge (HPGe) detector and corrected for decay, line intensity, self-shielding and photopeak efficiency to extract the number of $^{95,97}$Zr and $^{198}$Au products
(Supp. mat.). In the experiments, two different HPGe detectors (respectively shielded and unshielded) were used.

\begin{figure}[h]
\centering
      \begin{subfigure}[h]{0.7\columnwidth}
   \centering
      \includegraphics[width=0.7\columnwidth]{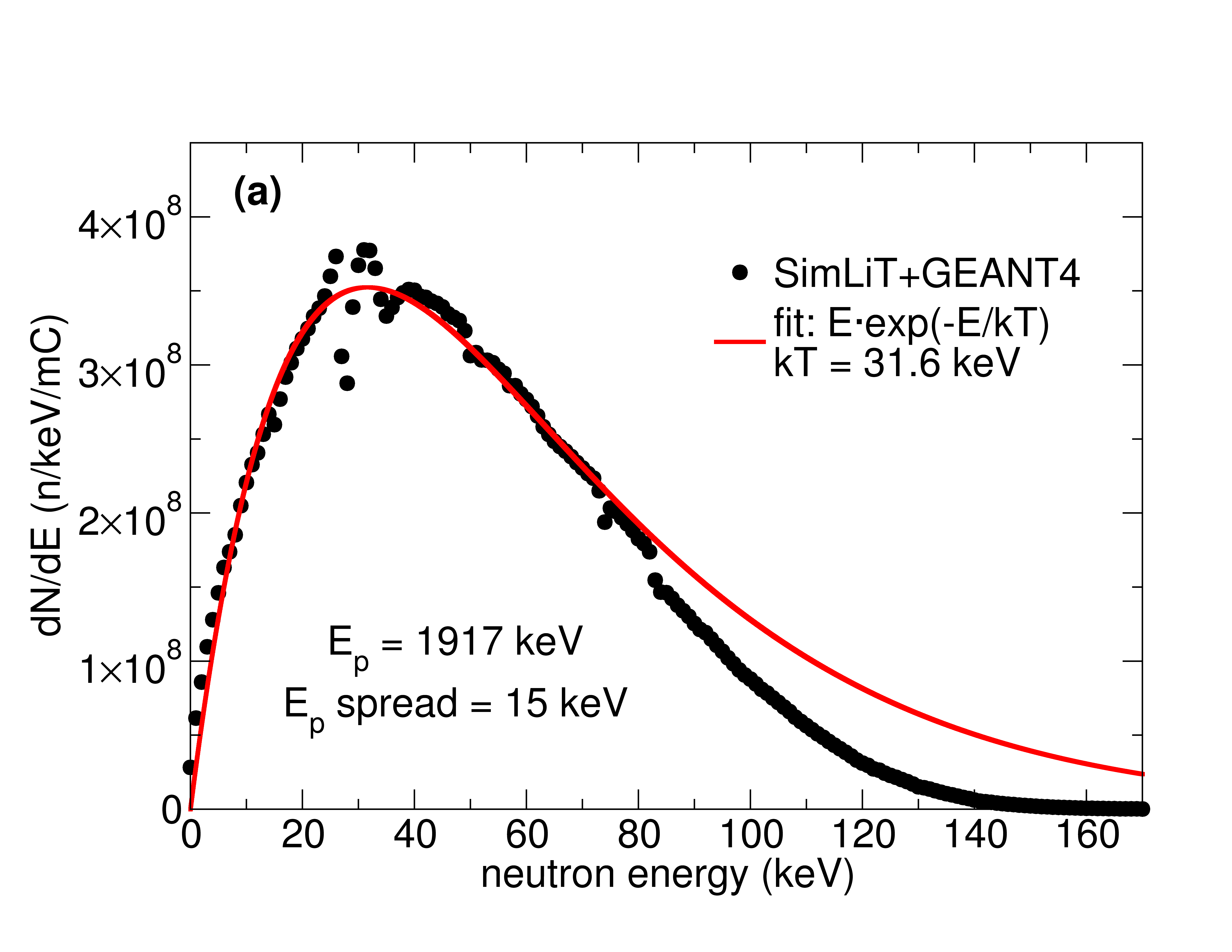}
   \end{subfigure}
\begin{subfigure}[b]{0.7\columnwidth}
\centering
   \includegraphics[width=0.7\columnwidth]{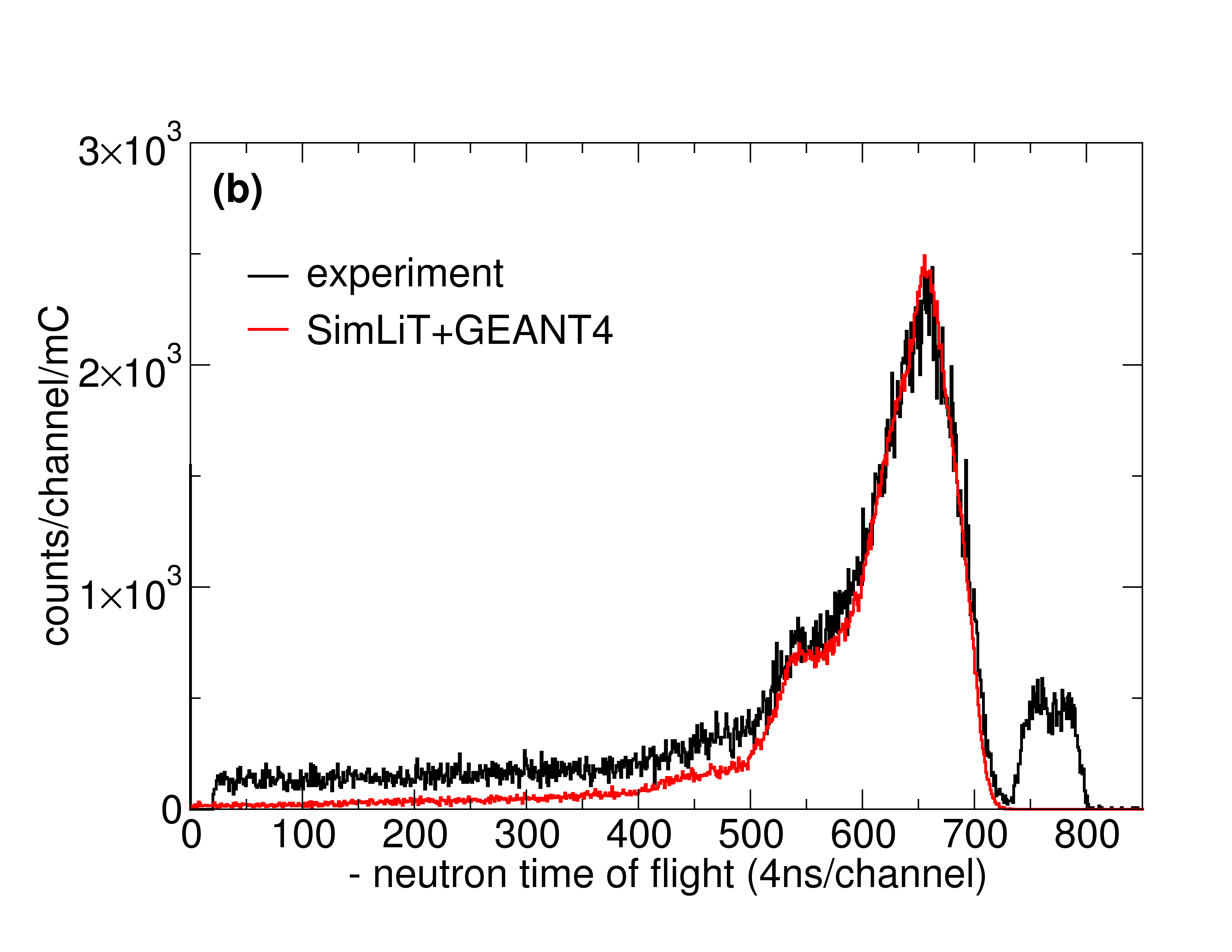}
\end{subfigure}
 \caption{(a) Simulated neutron spectrum $\left(\frac{dn_{sim}}{dE_n}\right)$ impinging on the Zr target (Exp. \Rmnum{2}) (dots) and fit (solid line) of a Maxwell-Boltzmann flux (see text).
 (b) neutron time-of-flight spectrum (black) measured with a 200 ns-wide chopped proton beam and a $1^{\prime \prime}$-thick $^6$Li-glass detector at a distance of 180 cm
 from the liquid-lithium target.
 The separated group at the right-end of the experimental spectrum results from prompt 478-keV $\gamma$ rays from $^7$Li$(p,p^{\prime})$.
 The spectrum obtained from a simulation of the entire experimental system using the SimLiT-GEANT4 codes is shown in red and reproduces closely the main neutron group.
 The neutron group around channel 550, observed in both the experimental and simulated spectrum is due to the n+$^{56}$Fe resonance at $E_n$ = 26 keV (see (a))
 and extensive neutron interactions in the stainless steel flanges between target and detector.
 The simulated spectrum, calculated in absolute time-of-flight units (ns), is converted into channels relative to the gamma peak starting at t=6 ns;
 its ordinate (counts/channel/mC) is calculated based on the proton charge accumulated in the measurement with no further normalization. Neutrons observed at large time-of-flight values are attributed to scattering off materials outside the scope of the simulation.}
\label{fig:spectrum}
\end{figure}

Characterization of the activation data in terms of a cross section requires knowledge of the neutron spectrum seen by the targets.
The integral neutron spectrum seen by a target under the conditions of the current experiment is however
not measurable and we rely for its shape on detailed simulations using the codes SimLiT \cite{SimLiT} for the thick-target $^7$Li$(p,n)$
neutron yield and GEANT4 \cite{GEANT4} for neutron transport (fig. \ref{fig:spectrum}a).
The SimLiT-GEANT4 simulations have been carefully benchmarked in a separate experiment \cite{Gitai} and excellent agreement with experimental time-of-flight and
(differential and integral) energy spectra was obtained \cite{Gitai,SimLiT}.
We also measured the neutron time-of-flight spectrum (fig. \ref{fig:spectrum}b) with the present LiLiT setup using a 1-inch thick $^6$Li-glass detector positioned at 0$\degree$ at a distance
of 180 cm downstream of the lithium target and a chopped proton beam (200 ns width). Despite extensive scattering of neutrons between source and detector,
the SimLiT-GEANT4 simulation is in good agreement with the measured spectrum and confirms its reliability.
The simulated spectrum $\frac{dn_{sim}}{dE_n}$ is well fitted in the range $E_n \sim 0-80$ keV by a Maxwell-Boltzmann (MB) flux
$v \frac{dn_{MB}}{dE_n} \propto E_n e^{-\frac{E_n}{kT}}$ with $kT\sim30$ keV (fig. \ref{fig:spectrum}a),
except for small glitches due to resonances of neutron reactions in structural materials (Fe, Al, Ni and Cr) having a negligible contribution to the integral yield.
The $^A$Zr$(n,\gamma)$ cross sections directly determined in the experiment (Table \ref{tab: MACS}), averaged over the neutron spectrum, are obtained from the expression
\begin{equation}
 \sigma_{exp}(A)=\sigma_{ENDF}(\textrm{Au})\frac{N_{act_{exp}}(A+1)}{N_{act_{exp}}(\textrm{Au})}\frac{n_{t}(\textrm{Au})}{n_{t}(A)}\frac{f(\textrm{Au})}{f(A)} \label{eq:sigma_i}
\end{equation}
where $N_{act_{exp}}(A+1)$ ($N_{act_{exp}}(\textrm{Au})$) is the number of $^{(A+1)}$Zr ($^{198}$Au) activated nuclei determined experimentally,
$n_t(A)$ ($n_t(\textrm{Au})$) is the $^A$Zr (Au) target thickness (atom/cm$^2$) and
$f(A)$ ($f(\textrm{Au})$) accounts for the decay of activated nuclei during the activation time and variations in the neutron rate (see Supp. mat.).
In (\ref{eq:sigma_i}), we use as reference a $^{197}$Au$(n,\gamma)$ cross section
\begin{equation}
 \sigma_{ENDF}(\textrm{Au})=\frac{\int\sigma_{ENDF}(E_n;\textrm{Au})\frac{dn_{sim}}{dE_n}dE_n}{\int\frac{dn_{sim}}{dE_n}dE_n}. \label{eq:sigma_endf}
\end{equation}
In (\ref{eq:sigma_endf}), $\sigma_{ENDF}(E_n;\textrm{Au})$ is taken from the ENDF/B-VII.1 (USA,2011) \cite{ENDF} library for $^{197}$Au.
The latter library (denoted henceforth ENDF) was extensively validated for $^{197}$Au \cite{Lederer, Gelina} and especially in the neutron energy range relevant to our
measurements \cite{Gitai,neutron_energies} and serves here for neutron fluence normalization.

\begin{figure}[h]
\centering
   \begin{subfigure}[h]{0.7\columnwidth}
   \centering
   \includegraphics[width=0.7\columnwidth]{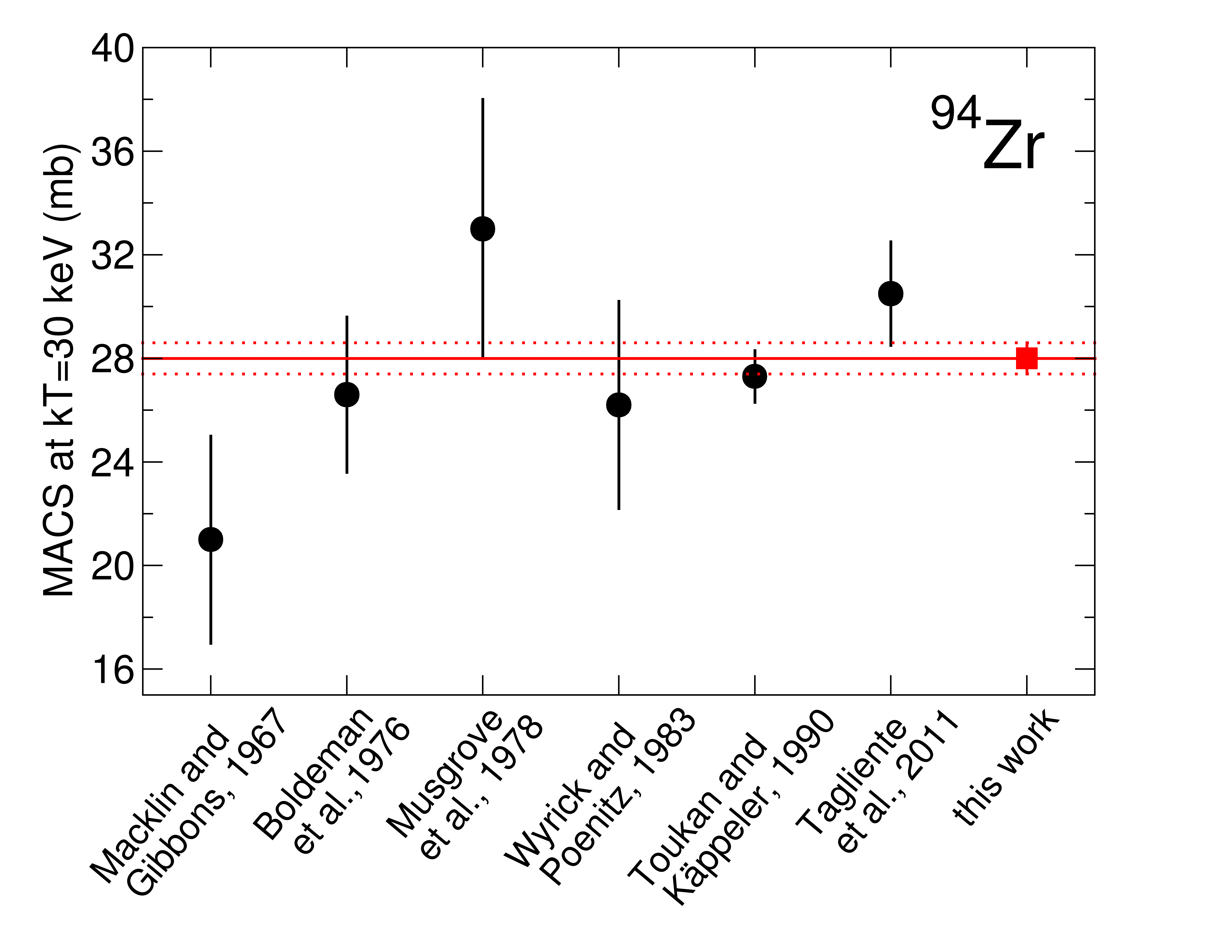}
\end{subfigure}
\begin{subfigure}[b]{0.7\columnwidth}
\centering
   \includegraphics[width=0.7\columnwidth]{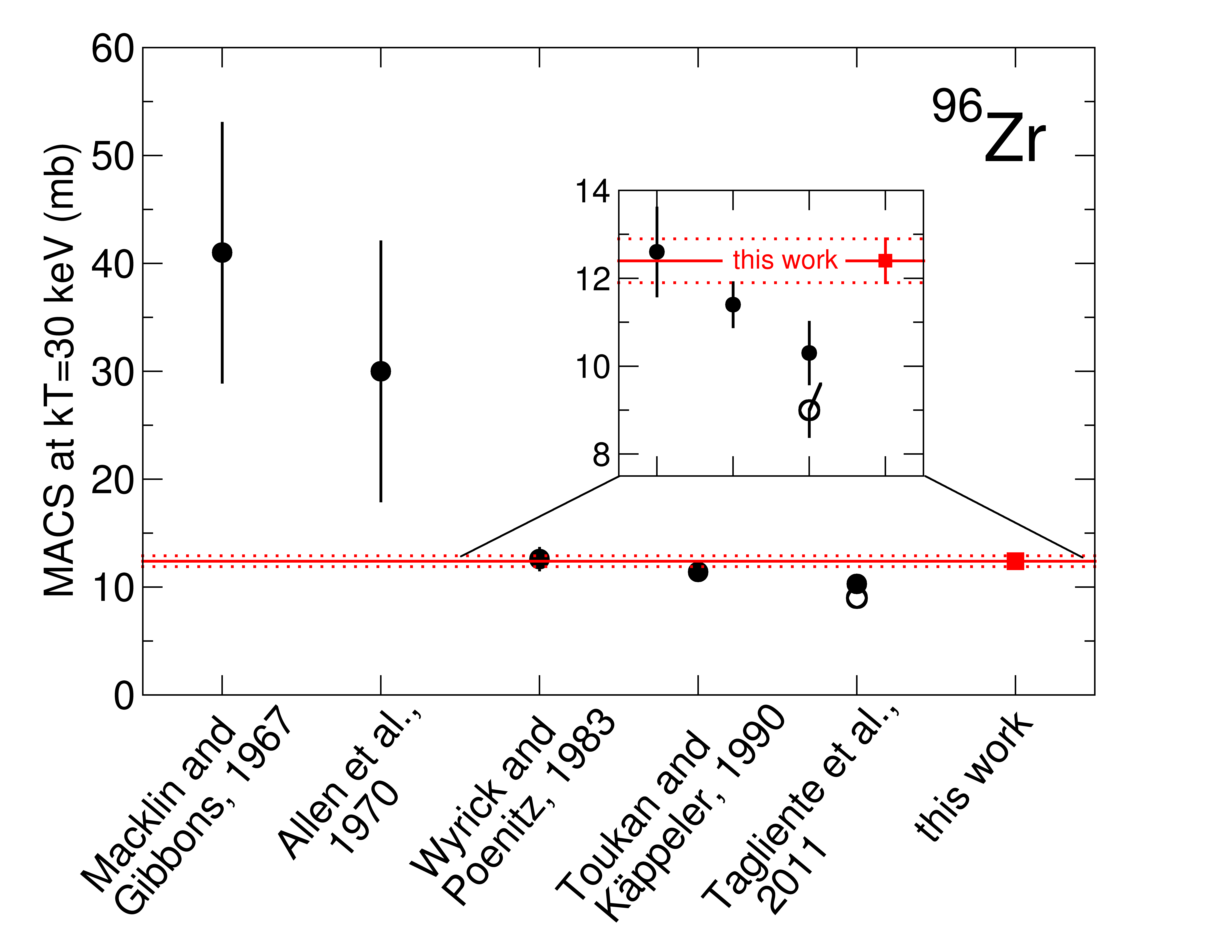}
\end{subfigure}
\caption{Comparison of MACS (30 keV) from the literature (black dots) \cite{Macklin,Boldeman,Musgrove,Wyrick,Tok90,Tagliente94,Allen,Tagliente96,Kadonis}
 and the present work (red squares) for $^{94}$Zr (top) and $^{96}$Zr (bottom). Wyrick et al. and Toukan et al. values \cite{Wyrick, Tok90}  obtained by activation using the $^{197}$Au$(n,\gamma)$
 cross section  as reference were corrected to the value established in \cite{Lederer,Gelina} as done in this work.
 For $^{96}$Zr, The open circle value is the value obtained by  by Tagliente et al. \cite{Tagliente96} from a time-of-flight measurement 
 and the full circle is the value obtained by adding a direct radiative capture (DRC) component.}
\label{fig:comp}
\end{figure}

In order to extrapolate the activation results and extract experimental MACS values we use available neutron cross section libraries, corrected by our activation data
in the measured energy range, and detailed SimLiT-GEANT4 simulations of the setup.
The MACS of a reaction at the temperature $T$ of an astrophysical site is defined as
\begin{equation}
 MACS(kT) = \frac{\Braket{\sigma v}}{v_{T}}=\frac{2}{\sqrt{\pi}}\frac{\intop_{0}^{\infty}\sigma(E_n)E_ne^{-\frac{E_n}{kT}}dE_n}{\intop_{0}^{\infty}E_ne^{-\frac{E_n}{kT}}dE_n} \label{eq:MACS_intro}
\end{equation}
where $\sigma(E_n)$ is the true energy-dependent reaction cross section. We determine here an experimental value MACS$_{lib}^{exp}$ of the radiative neutron capture
$(n,\gamma)$ for the target nucleus $^A$Zr (A = 94,96) using the expression:
\begin{equation}
 \textrm{MACS}_{lib}^{exp}(A;kT) = \frac{2}{\sqrt{\pi}}\frac{\intop_{0}^{\infty} C_{lib}(A)\sigma_{lib}^A(E_{n})E_{n}e^{-\frac{E_{n}}{kT}}dE_{n}}{\intop_{0}^{\infty}E_{n}e^{-\frac{E_{n}}{kT}}dE_{n}}. \label{eq:MACS_C}
\end{equation}
In (\ref{eq:MACS_C}), $\sigma_{lib}^A(E_n)$ is the $^A$Zr$(n,\gamma)$ cross section given by a neutron library $lib$ and $C_{lib}(A)$ a correction factor for the library $lib$
extracted from our activity measurements by the expression
\begin{equation}
  C_{lib}\left(A\right)=\frac{\left(\frac{N_{act_{exp}}\left(A+1\right)}{N_{act_{exp}}\left(\textrm{Au}\right)}\right)}{\left(\frac{N_{act_{lib}}\left(A+1\right)}{N_{act_{ENDF}}\left(\textrm{Au}\right)}\right)}. \label{eq:C}
\end{equation}
In (\ref{eq:C}), $N_{act_{lib}}(A+1)$ and $N_{act_{ENDF}}$(Au) are the number of activated $^{A+1}$Zr and $^{198}$Au nuclei obtained in a single overall SimLiT-GEANT4 simulation
of the experimental system (including the Au-Zr target sandwich), using the neutron library $lib$ for $^A$Zr. The final results for the MACS (30 keV) of $^{94,96}$Zr,
listed in Table \ref{tab: MACS}, are extracted using the ENDF library (see Supp. mat. Table 11 for a comparison of MACS values using extrapolation with different neutron libraries).
\begin{table*}
\centering
\caption{Experimental parameters ($E_p$: proton mean energy and $\Delta E_p$: energy spread (1$\sigma$), $\Delta z$: distance lithium surface-activation target and accumulated proton charge
during the activations), experimental cross sections (multiplied by $\frac{2}{\sqrt{\pi}}$) and MACS (30 keV) values determined in this work for $^{94,96}$Zr (see text and Supp. mat.).
The final value of the MACS is obtained by an unweighted average of the three experiments and the uncertainty is determined based on the individual uncertainties taking into consideration
their systematic component.}
\label{tab: MACS}
\ra{1.3}
\begin{tabular}{l c c c c c c c}
\toprule
Exp&$E_p\pm\Delta E_p$ & $\Delta z$ & charge      &Isotope & $\frac{2}{\sqrt{\pi}}\sigma_{exp}$ & $C_{ENDF}$ & MACS$_{ENDF}^{exp}$ \\ 
   &(keV) &    (mm)    & (mA$\cdot$h)&        &              (mb)                        &           &  (mb) \\[0.5ex]
\midrule
\multirow{2}{*}{\Rmnum{1}}&\multirow{2}{*}{$1908\pm15$}&\multirow{2}{*}{8}&\multirow{2}{*}{1.1}&$^{94}$Zr & $29.9\pm0.8$ & $0.98\pm0.03$ & $28.7\pm0.8$\\
&&&&$^{96}$Zr & $13.6\pm0.7$ & $1.22\pm0.08$ &$12.6\pm0.8$\\
\midrule
\multirow{2}{*}{\Rmnum{2}}&\multirow{2}{*}{$1917\pm15$}&\multirow{2}{*}{6}&\multirow{2}{*}{1.1}&$^{94}$Zr & $28.2\pm0.8$ & $0.96\pm0.03$ & $27.9\pm0.8$\\
&&&&$^{96}$Zr & $13.4\pm0.5$ & $1.18\pm0.04$ & $12.1\pm0.5$\\
\midrule
\multirow{2}{*}{\Rmnum{3}}&\multirow{2}{*}{$1942\pm15$}&\multirow{2}{*}{6}&\multirow{2}{*}{1.5}&$^{94}$Zr & $24.5\pm0.7$ & $0.93\pm0.03$ & $27.3\pm0.7$\\
&&&&$^{96}$Zr & $12.3\pm0.3$ & $1.20\pm0.06$ & $12.4\pm0.6$\\
\midrule
unweighted&&&&$^{94}$Zr &&& $28.0\pm0.6$ \\
average&&&&$^{96}$Zr &&& $12.4\pm0.5$ \\ [1ex]
\bottomrule
\end{tabular}
\end{table*}
The values $\frac{2}{\sqrt{\pi}}\cdot \sigma_{exp}$
(which depends on the proton incident energy via the resulting neutron spectrum) and the MACS (a property of the nuclide) differ by 4\% to 13 \%, giving a measure of the moderate
correction involved in the extrapolation to the MACS. Table \ref{table: uncertainties1} (and Supp. mat.) lists the uncertainties in the MACS values determined in one of our experiments.
In order to have a quantitative estimate of the uncertainty associated with the use of a simulated neutron energy spectrum, we use the data of \cite{Gitai} as follows.
The $^7$Li$(p,n)$ neutron time-of-flight spectra measured in \cite{Gitai} in the range $0\degree-80\degree$ and those simulated by SimLiT-GEANT4 in the conditions
of this experiment were converted to energy spectra using the same algorithm (see \cite{Gitai,SimLiT} for details of algorithm) and the two resulting spectra were then
convoluted with the same energy-dependent $^{197}$Au$(n,\gamma)$ ENDF cross section. The resulting averaged cross sections are 608 mb and 599 mb, respectively and an
uncertainty of 1.5\% is correspondingly ascribed to the use of the simulated spectrum for cross section calculation.
An uncertainty component resulting from the proton energy $E_p$ and energy spread $\Delta E_p$ was estimated from the change of MACS values when distributing
$E_p$ and $\Delta E_p$ in their respective range; we note the insensitivity of the final value to $E_p$ and $\Delta E_p$.
The uncertainty associated with the use of the ENDF library for the extrapolation to the MB spectrum was calculated (see Supp. mat. for details) based on the quoted ENDF (energy-dependent)
uncertainties for Au and $^{94,96}$Zr \cite{ENDF_Unc}.
\begin{table}[h]
\centering
\caption{Random (rand) and systematic (sys) relative uncertainties in the MACS (30 keV) of $^{94,96}$Zr for Exp. \Rmnum{2} (see text and Supp. mat.).}
\label{table: uncertainties1}
\ra{1.3}
\centerline{
\begin{tabular}{l c c c c c c }
\toprule
 & \multicolumn{5}{c}{Uncertainty (\%)}\\
\cmidrule {2-6}
 & \multicolumn{2}{c}{$^{94}$Zr} && \multicolumn{2}{c}{$^{96}$Zr}\\
 Source of uncertainty&rand&sys&&rand&sys\\ [0.5ex]
 \midrule
target thickness&  0.4&&& 0.4\\
activated nuclei &  1.6&&& 0.6 \\
photopeak eff. rel. to Au && 0.5 &&& 0.5 \\
simulation&&  1.5&&& 1.5\\
E$_p$, $\Delta$E$_p$ and $\Delta$z&& 0.4&&& 2.3\\
$\sigma_{ENDF}$(Au)&&1.0&&&1.0\\
$\sigma_{ENDF}$(Zr)&&1.6&&&2.2\\
\midrule
Total random uncertainty&1.6&&& 0.7\\
Total systematic uncertainty&&2.5 &&&3.7\\
\midrule
Total uncertainty &\multicolumn{2}{c}{3.0} && \multicolumn{2}{c}{3.8}\\ [1ex]
\bottomrule
\end{tabular}}
\end{table}
Fig. \ref{fig:comp} illustrates a comparison of our results with existing sets of experimental data for $^{94,96}$Zr MACS (30 keV) values.
For measurements obtained by activation \cite{Wyrick, Tok90}, values in fig. \ref{fig:comp} were corrected for sake of consistency to the ENDF $^{197}$Au$(n,\gamma)$ cross section
used in the present analysis and established since as reference value \cite{Lederer, Gelina} and for updated photo-peak intensities used in this work \cite{Datasheet198,Datasheet97} (see Table 4 in Supp. mat.).
We stress the lower uncertainties compared to most experiments, owed to both the higher neutron intensity
and corresponding better counting statistics and the detailed simulations of the experimental setup.
We observe in general a slightly larger uncertainty for $^{96}$Zr (and also larger correction factors $C_{lib}$(96)).
We expect the present results to be significant in $s$-process calculations in the Zr region and we note especially that
the lower $^{96}$Zr$(n,\gamma)^{97}$Zr MACS value (open circle in Fig. \ref{fig:comp}) used recently in the detailed astrophysical model calculations by Lugaro et al. \cite{Lugaro} is inconsistent
with our result. This lower value was in fact corrected in \cite{Tagliente96} by adding a direct-capture component; the corrected value (full circle in Fig. \ref{fig:comp}) is consistent with the present work.
Although the thermal energy of 30 keV considered so far is widely adopted as a reference point for $s$-process nucleosynthesis, the relevant values for its ``weak'' (``main'') regimes are considered
to be 90 keV (8 and 23 keV) \cite{s_proc}. Since Zr lies at the border line between these regimes, we extrapolate our experimental values for $^{94,96}$Zr to these
temperatures (see Supp. mat. for details of these extrapolations).
From our analysis, the MACS at 8, 23 and 90 keV for $^{94}$Zr ($^{96}$Zr) are $63.9\pm3.7$ ($45.3\pm5.5$), $32.5\pm1.1$ ($15.6\pm1.1$) and $18.5\pm2.1$ ($8.6\pm1.7$) respectively.
The larger uncertainties are due mainly to the smaller overlap between the experimental neutron spectrum and the respective MB spectra at temperatures far from 
30 keV; they are based here also on the quoted ENDF energy-dependent uncertainties $\Delta \sigma_{ENDF}$ (Supp. mat.). Any future update in these uncertainties could be 
applied using the same formalism.

An additional feature of the use of a thick liquid-lithium target with a high-intensity proton beam is the copious production of high-energy $\gamma$ rays from the
radiative capture $^7$Li$(p,\gamma)^8$Be reaction. We observe these $\gamma$ rays in our experiments via the photonuclear reaction $^{90}$Zr$(\gamma,n)^{89}$Zr
(fig. \ref{fig:gamma_spec_old}). No other $(\gamma,n)$ reaction on the $^{nat}$Zr target is readily observable by $\gamma$ spectrometry of the activated target;
we note also that the $^{96}$Zr$(\gamma,n)^{95}$Zr reaction (which could potentially interfere with the $^{94}$Zr$(n,\gamma)^{95}$Zr activation)
has negligible yield compared to that of the $(n,\gamma)$ reaction due to the low $^{96}$Zr abundance.
The $^7$Li$(p,\gamma)^8$Be reaction produces principally 17.6 MeV and 14.6 MeV $\gamma$ rays and their yield was measured in \cite{7Li_p_g_1} with a thin Li target.
The high-energy $\gamma$ spectrum was measured (Supp. mat.) with the LiLiT setup in a separate experiment (under neutron threshold) with a $6^{\prime\prime}\times4^{\prime\prime}$ NaI(Tl) detector
positioned behind a 1.5 m thick concrete wall (shielding the overwhelming 478-keV $\gamma$-rays from
$^7$Li$(p,p^ {\prime}\gamma)$). Using $^{90}$Zr$(\gamma,n)^{89}$Zr cross section values of 173 mb (85 mb) for $E_{\gamma_{0}} = 17.6$ ($E_{\gamma_{1}} = 14.6$) MeV
(in the Giant Dipole Resonance region) measured in \cite{Atlas} and an averaged branching ratio $\frac{\gamma_{1}}{\gamma_{0}} \sim 1.3$
obtained by integrating the data of Zahnow et al. \cite{7Li_p_g_1} into a thick-target yield between 0.1 $\leq$ $E_p$ $\leq$ 1.9 MeV,
the respective measured gamma yields are 3$\times10^8$ $\gamma_0$/s/mA (4$\times10^8$ $\gamma_1$/s/mA).
These yields can be compared with  the values 1.2$\times10^8$ $\gamma_0$/s/mA (1.5$\times10^8$ $\gamma_1$/s/mA) calculated from the
data of \cite{7Li_p_g_1} and show a considerable additional yield, possibly due in part to additional resonances in $^8$Be for $E_p > 1.5$ MeV, above the range measured in \cite{7Li_p_g_1}.

In conclusion, we have shown that the high-power Liquid-Lithium Target bombarded by a mA proton beam, in conjunction with detailed simulations of the experimental system,
allows us precise determination of 30-keV MACS values. In this first experiment, we determined the 30-keV MACS of $^{94}$Zr and $^{96}$Zr as $28.0\pm0.6$ mb and $12.4\pm0.5$ mb,
respectively.
The SARAF-LiLiT facility is being upgraded in several aspects. Since the LiLiT device is capable of sustaining higher power levels than those used
so far, the primary proton intensity, presently on average of $\sim$ 1 mA, is being upgraded to 2 mA.
The neutron intensity, intersecting a small target (6 mm diameter) at a typical distance of 5.5 mm from the liquid Li surface, is estimated as
$\sim3\times10^{10}$ n/s for a 2-mA proton beam at 1.93 MeV, considered suitable in view of the energy spread.
It is expected that, with gained experience on operation and control of the accelerator, the beam limitation 
will be improved towards a final goal of 4 mA.
A pneumatic rabbit is in construction for the transport (\textit{in vacuo}) of activation targets with short half-life products
(down to a few seconds). Finally a dedicated target room, designed for the housing of an upgraded LiLiT-II neutron source,
will give more flexibility in the use of the facility.
The system will be particularly useful for neutron activation measurements of low-abundance isotopes or radioactive targets.
As an example, a measurement of the important $^{60}$Fe$(n,\gamma)^{61}$Fe cross section, as performed in \cite{60Fe}, would be possible with a
$\sim$100 ng $^{60}$Fe radioactive target or correspondingly, improve the statistical uncertainty with a target of larger mass.
We stress also the possibility to use the present setup (above or below neutron threshold) for the study of photonuclear reactions in the Giant Dipole Resonance region
with intense yields of $\sim7\times10^8$ $\gamma/$s/mA from prompt $^7$Li$(p,\gamma)$ capture $\gamma$ rays, as demonstrated here with the $^{90}$Zr$(\gamma,n)^{89}$Zr
photonuclear reaction. The setup is planned to be used towards investigations of photodissociation and photofission reactions.

We acknowledge gratefully the support of the Pazi Foundation (Israel) and of the German-Israeli Foundation (GIF Research Grant No. G-1051-103.7/2009).


\newpage

\begin{center}
\textbf{{\huge Supplementary material}}
\end{center}
\setcounter{equation}{5}
\setcounter{table}{2}
\setcounter{figure}{4}

\vspace{5ex}
\section{Activation target properties}
\begin{table}[h]
\centering
\caption{Properties of $^{nat}$Zr and $^{197}$Au targets used in the LiLiT experiments.
The order of the samples listed for each experiment is the order in which the samples were placed downstream the Li target.}
\label{table: samples}
\ra{1.3}
\begin{tabular}{l c c c c}
\toprule
Exp &Sample & Diameter (mm) & Mass (mg) & $n_t$ ($cm^{-2}$)\\ [0.5ex]
\midrule
\multirow{3}{*}{\Rmnum{1}}&Au-1 & 24 & 184.0(1) & $1.244(4)\times10^{20}$\\
&$^{nat}$Zr-1 & 24 & 60.80(1) & $8.87(3)\times10^{19}$\\
&Au-2 & 24 & 185.7(1) & $1.255(5)\times10^{20}$\\
\midrule
\multirow{3}{*}{\Rmnum{2}}&Au-6 & 25 & 115.85(7) & $7.22(2)\times10^{19}$\\
&$^{nat}$Zr-X & 25 & 312.5(1) & $4.20(1)\times10^{20}$\\
&Au-7p & 25 & 116.73(6) & $7.27(2)\times10^{19}$\\
\midrule
\multirow{3}{*}{\Rmnum{3}}&Au-16 & 25 & 116.93(6) & $7.28(2)\times10^{19}$\\
&$^{nat}$Zr-3 & 25 & 312.10(6) & $4.20(1)\times10^{20}$\\
&Au-17 & 25 & 112.30(6) & $6.84(2)\times10^{19}$\\ [1ex]
\bottomrule
\end{tabular}
\end{table}
\FloatBarrier

\section{Activation results}
The number of activated nuclei at the end of the irradiation, $N_{act_{exp}}$, is obtained from equation (\ref{eq:N_act_exp})
\begin{equation}
 N_{act_{exp}}=\frac{C}{\epsilon_{\gamma}I_{\gamma}K_{\gamma}}\frac{e^{\lambda t_{cool}}}{1-e^{-\lambda t_{real}}}\frac{t_{real}}{t_{live}}, \label{eq:N_act_exp}
\end{equation}
where $C$ is the number of counts in the photo-peak, $\epsilon_{\gamma}$ is the detector photoelectric efficiency, $I_{\gamma}$ is the $\gamma$-intensity per decay,
$K_{\gamma}$ is the correction due to $\gamma$-ray self absorption in the sample. In the case of a disk sample of thickness $x$, $K_{\gamma}=\frac{1-e^{-\mu x}}{\mu x}$
where $\mu$ is the $\gamma$-ray absorption coefficient \cite{K_gamma}. 
 $\lambda=\frac{ln(2)}{t_{\frac{1}{2}}}$ is the decay constant of the activated nucleus, $t_{cool}$ is the time between the end of the irradiation and the activity measurement,
$t_{real}$ is the real measuring time, and $t_{live}$ is the live measuring time. The decay parameters used in this analysis are listed in Table \ref{table: decay}.
The $\gamma$-ray absorption coefficients, $\mu$, used in this analysis were taken from \cite{NIST}.
\begin{table}[h]
\centering
\caption{Properties of the relevant target and product nuclei used in this work.}
\label{table: decay}
\ra{1.3}
\begin{tabular}{l c c c c c c}
\toprule
Target& natural& Product & Half-life, & $\gamma$-ray energy,& Intensity per & Ref.\\
nucleus& abundance& nucleus &  $t_{\frac{1}{2}}$ & $E_{\gamma}$ (keV) &  decay, $I_{\gamma}$ (\%) \\ [0.5ex]
\midrule
$^{197}$Au& 1& $^{198}$Au & $2.6947(3)$ d & $411.80205(17)$ & $95.62(6)$ & \cite{Datasheet198}\\
$^{90}$Zr & 0.5145 & $^{89}$Zr & $78.41(12)$ h & $909.15(15)$ & $99.04(3)$& \cite{Datasheet89}\\ 
$^{94}$Zr & 0.1738&$^{95}$Zr & $64.032(6)$ d &$756.725(12)$ & $54.38(22)$ & \cite{Datasheet95}\\
$^{96}$Zr & 0.028& $^{97}$Zr & $16.749(8)$ h &$743.36(3)$ & $93.09(16)$ & \cite{Datasheet97}\\ [1ex]
\bottomrule
\end{tabular}
\end{table}
\FloatBarrier
The number of activated nuclei at the end of the irradiation, $N_{act_{exp}}$, obtained from equation (\ref{eq:N_act_exp}) are summarized in Table \ref{table: Nact summary}.

\begin{table}[h]
\centering
\caption{Summary of the number of activated nuclei at the end of the three experiments, $N_{act_{exp}}$. $E_p$ is the laboratory proton beam mean energy, $\Delta E_p$ the energy spread (1$\sigma$), and $\Delta$z is the Li to target distance.
$f(A)$, $f(Au)$ is a correction applied to $N_{act_{exp}}$ to account for the decay of activated nuclei during the activation
time ($t_a$) and the variations in the neutron rate: 
$f(A)=\frac{\intop_{0}^{t_{a}}\Phi(t)e^{-\lambda_{A}(t_{a}-t)}dt}{\intop_{0}^{t_{a}}\Phi(t)dt}$
(for a constant neutron flux, $f(A)$ reduces to $f(A)=\frac{1-e^{-\lambda_A t_a}}{\lambda_A t_a})$.}
\label{table: Nact summary}
\ra{1.3}
\begin{tabular}{l c c c c c c c}
\toprule
Exp&$E_p\pm\Delta E_p$ (keV)&$\Delta z$ (mm)&Sample& Nuclei & $N_{act_{exp}}$&$f(A)$, $f(Au)$\\ [0.5ex]
\midrule
 \multirow{4}{*}{\Rmnum{1}}&\multirow{4}{*}{$1908\pm15$}&\multirow{4}{*}{8}&Au-1& $^{198}$Au& $6.78\pm0.08\times 10^9$& 0.99\\
 &&&\multirow{2}{*}{$^{nat}$Zr-1} & $^{95}$Zr & $3.83\pm0.05\times10^7$& 1.0\\
 &&&& $^{97}$Zr & $2.72\pm0.03\times10^6$&0.96\\
 &&&Au-2 & $^{198}$Au & $6.73\pm0.08\times10^9$&0.99\\
\midrule
\multirow{4}{*}{\Rmnum{2}}&\multirow{4}{*}{$1917\pm15$}&\multirow{4}{*}{6}&Au-6 & $^{198}$Au & $5.4\pm0.1\times10^9$&0.99\\
  &&&\multirow{2}{*}{$^{nat}$Zr-X} & $^{95}$Zr & $2.34\pm0.06\times10^8$& 1.0\\
 &&&& $^{97}$Zr & $1.75\pm0.03\times10^7$& 0.98\\
 &&&Au-7p & $^{198}$Au & $5.3\pm0.1\times10^9$&0.99\\
 \midrule
\multirow{4}{*}{\Rmnum{3}}&\multirow{4}{*}{$1942\pm15$}&\multirow{4}{*}{6}&Au-16 & $^{198}$Au & $1.26\pm0.03\times10^{10}$& 0.99\\
  &&&\multirow{2}{*}{$^{nat}$Zr-3} & $^{95}$Zr & $5.3\pm0.1\times10^8$&1.0\\
 &&&& $^{97}$Zr & $4.18\pm0.08\times10^7$& 0.98\\
 &&&Au-17 & $^{198}$Au & $1.18\pm0.03\times10^{10}$&0.99\\ [1ex]
\bottomrule
\end{tabular}
\end{table}
\FloatBarrier

\section{Experimental cross section \label{sec:exp_CS}}
\begin{table}[h]
\centering
\caption{Summary of experimental and ENDF library averaged cross section results. $\sigma_{exp}$ is calculated from Eq. (1), $\sigma_{ENDF}$ is averaged over the simulated
neutron spectrum $\frac{dn_{sim}}{dE_n}$. $E_p$ is the proton beam mean energy and $\Delta E_p$ the energy spread (1$\sigma$), $\Delta z$ the distance lithium surface - activation target.}
\label{table: CS summary}
\ra{1.3}
\begin{tabular}{l c c c c c}
\toprule
Exp&$E_p\pm\Delta E_p$ (keV)&$\Delta z$ (mm)& Isotope & $\sigma_{exp}$ (mb) & $\sigma_{ENDF}$ (mb)\\ [0.5ex]
\midrule
\multirow{3}{*}{\Rmnum{1}}&\multirow{3}{*}{$1908\pm15$}&\multirow{3}{*}{8}&$^{197}$Au &  &$583\pm12$ \\
&&&$^{94}$Zr &$26.5\pm0.7$ &$26.8\pm0.5$ \\
&&&$^{96}$Zr &$12.1\pm0.6$ &$9.5\pm0.2$ \\
\midrule
\multirow{3}{*}{\Rmnum{2}}&\multirow{3}{*}{$1917\pm15$}&\multirow{3}{*}{6}&$^{197}$Au &  &$580\pm12$ \\
&&&$^{94}$Zr &$25.0\pm0.7$ &$26.4\pm0.5$ \\
&&&$^{96}$Zr &$11.9\pm0.4$ &$9.9\pm0.2$ \\ 
\midrule
\multirow{3}{*}{\Rmnum{3}}&\multirow{3}{*}{$1942\pm15$}&\multirow{3}{*}{6}&$^{197}$Au &  &$521\pm10$ \\
&&&$^{94}$Zr &$21.7\pm0.6$ &$23.9\pm0.5$ \\
&&&$^{96}$Zr &$10.9\pm0.3$ &$9.1\pm0.2$ \\ [1ex]
\bottomrule
\end{tabular}
\end{table}
\FloatBarrier

\section{\label{sec: unc.}Uncertainties}
\subsection{Beam and target parameters}
\begin{table}[h]
\centering
\caption{Sensitivity of extracted MACS in Exp. \Rmnum{2} to changes in proton beam energy within its uncertainty range for an energy spread of 15 keV.}
\label{table: proton energy change}
\ra{1.3}
\begin{tabular}{c c c}
\toprule
 $E_p$ (keV)& \multicolumn{2}{c}{MACS (mb)}\\
& $^{94}$Zr & $^{96}$Zr\\ [0.5ex]
\midrule
1914&27.9&12.1\\
1915&27.9&12.1\\
1916&27.9&12.2\\
\textbf{1917}&\textbf{27.9}&\textbf{12.1}\\
1918&27.9&12.1\\
1919&28.0&12.0\\
1920&28.0&11.9\\
\midrule
standard deviation&0.05&0.1\\[1ex]
\bottomrule
\end{tabular}
\end{table}
\FloatBarrier

\begin{table}[h]
\centering
\caption{Sensitivity of extracted MACS in Exp. \Rmnum{2} to changes in proton beam energy spread within its uncertainty range for E$_p$=1915 keV.}
\label{table: proton energy spread change}
\ra{1.3}
\begin{tabular}{c c c}
\toprule
$\Delta$E$_p$ (keV)&\multicolumn{2}{c}{MACS (mb)}\\
& $^{94}$Zr & $^{96}$Zr\\ [0.5ex]
\midrule
5&28.0&11.8\\
10&28.0&11.9\\
\textbf{15}&\textbf{27.9}&\textbf{12.1}\\
20&27.8&12.2\\
25&27.9&12.2\\
\midrule
standard deviation&0.1&0.2\\[1ex]
\bottomrule
\end{tabular}
\end{table}
\FloatBarrier

\begin{table}[h]
\centering
\caption{Sensitivity of extracted MACS in Exp. \Rmnum{2} to changes in target distance from lithium surface within its uncertainty range for E$_p$=1915 keV and an energy spread of 15 keV.}
\label{table: distance change}
\ra{1.3}
\begin{tabular}{c c c}
\toprule
$\Delta$z (mm)&\multicolumn{2}{c}{MACS (mb)}\\
& $^{94}$Zr & $^{96}$Zr\\ [0.5ex]
\midrule
5&27.9&11.9\\
\textbf{6}&\textbf{27.9}&\textbf{12.1}\\
7&28.0&12.3\\
\midrule
standard deviation&0.1&0.2\\[1ex]
\bottomrule
\end{tabular}
\end{table}
\FloatBarrier

\subsection{Extrapolation to MB distribution}
In order to separate the contributions to the MACS uncertainty resulting from the experimentally measured value $\sigma_{exp}$
and from the extrapolation to a MB distribution, we express the MACS (30 keV) as:
\begin{equation}
 \textrm{MACS(30{\ }keV)} = \frac{2}{\sqrt{\pi}}\intop_{0}^{\infty}\phi_{MB}(30)\sigma dE = \frac{2}{\sqrt{\pi}}\left(\intop_{0}^{\infty}\phi_{exp} \sigma dE +
 \intop_{0}^{\infty}\left(\phi_{MB}(30) - \phi_{exp}\right) \sigma dE\right) \label{eq:MACS_30}
\end{equation}
where $\phi_{MB}(30)$ and $\phi_{exp}$ are the normalized MB and experimental neutron energy distribution respectively and $\sigma$ is the true energy dependent cross section.
The first term in Eq. (\ref{eq:MACS_30}) is 
\begin{equation}
 \sigma_{exp} = \intop_{0}^{\infty}\phi_{exp}\sigma dE = C_{ENDF}\intop_{0}^{\infty}\phi_{exp}\sigma_{ENDF} dE
\end{equation}
(Eq. (1) and (5)) and includes the
experimental uncertainties listed in Table \ref{table: CS summary}.
The second term in Eq. (\ref{eq:MACS_30}) can be re-written as
\begin{equation}
 \intop_{0}^{\infty}\left(\phi_{MB}(30) - \phi_{exp}\right) \sigma dE = C_{ENDF}\intop_{0}^{\infty}\left(\phi_{MB}(30) - \phi_{exp}\right) \sigma_{ENDF} dE
\end{equation}
and contributes to the uncertainty to the extent that $C_{ENDF}$ determined in the experimental energy range ($\phi_{exp}$) is not valid in the whole MB energy range due to
uncertainties $\Delta\sigma_{ENDF}$. We obtain thus the overall uncertainty as
\begin{equation}
 \Delta \textrm{MACS (30{\ } keV)} = \frac{2}{\sqrt{\pi}}\sqrt{\Delta\sigma_{exp}^2 + \left(C_{ENDF}\intop_{0}^{\infty}\left(\phi_{MB}(30) - \phi_{exp}\right) \Delta\sigma_{ENDF} dE \right)^2}\label{eq:delta_MACS_30}
\end{equation}
where $\Delta \sigma_{ENDF}$ are taken from the reported uncertainties for the ENDF cross section library \cite{ENDF_Unc}.
Note that we conservatively use here the quoted $\Delta \sigma_{ENDF}$ although an overall correction ($C_{ENDF}$) was already applied and that $\Delta \sigma_{ENDF}$
in different energy bins are added linearly and not in quadrature.
The values of $\Delta \textrm{MACS (30{\ } keV)}$ are listed in Tables \ref{table: uncertainties1} and \ref{table: uncertainties2}.

\subsection{Extrapolation from 30 keV to $kT$ = 8, 23 and 90 keV}
Similarly to Eq. (\ref{eq:MACS_30}), The MACS at any energy $kT$ can be denoted as
\begin{equation}
 \textrm{MACS}(kT) = \frac{2}{\sqrt{\pi}}\intop_{0}^{\infty}\phi_{MB}(kT)\sigma dE = \frac{2}{\sqrt{\pi}}\left(\intop_{0}^{\infty}\phi_{MB}(30)\sigma dE + \intop_{0}^{\infty}\left(\phi_{MB}(kT) - \phi_{MB}(30)\right)\sigma dE\right). \label{eq:MACS_kT}
\end{equation}
The first term is the MACS at 30 keV, so equation (\ref{eq:MACS_kT}) can be written as
\begin{equation}
 \textrm{MACS}(kT) = \textrm{MACS(30{\ }keV)} + \frac{2}{\sqrt{\pi}}\intop_{0}^{\infty}\left(\phi_{MB}(kT) - \phi_{MB}(30)\right)C_{ENDF}\sigma_{ENDF} dE. \label{eq:MACS_kT_2}
\end{equation}
The uncertainty is therefore given by
\begin{equation}
 \Delta \textrm{MACS}(kT) = \sqrt{\left(\Delta \textrm{MACS (30{\ } keV)}\right)^2 + \left( \frac{2}{\sqrt{\pi}}\intop_{0}^{\infty}\left(\phi_{MB}(kT) - \phi_{MB}(30)\right)C_{ENDF}\Delta\sigma_{ENDF} dE \right)^2}. \label{eq:delta_MACS_kT}
\end{equation}
Here again as in Eq. (\ref{eq:delta_MACS_30}), the second term under the square root is an estimate of the uncertainty due to the fact that the correction $C_{lib}$, valid in the experimental range
(quasi-MB at 30 keV), may not be valid in a different energy range. This uncertainty estimate is based on the ENDF library uncertainties \cite{ENDF_Unc} in the corresponding energy range.
It is also useful to calculate the $\textrm{MACS}(kT)$ using other neutron libraries \cite{ENDF, Cendl, Jendl, Jeff, Rosfond} with Eq. (\ref{eq:MACS_kT_2}), see Tables \ref{table: MACS_comp} and \ref{table: MACS_comp_2}.
Except for ENDF, the absence of quoted library uncertainties prevents calculating an associated MACS uncertainty. We note the close agreement between MACS values from the different (corrected)
libraries at 30 and 23 keV and the expected larger spread of values for 8 and 90 keV. We adopt as our final values for MACS at 8, 23 and 90 keV the average values of the different libraries and as
the uncertainties the larger between Eq. (\ref{eq:delta_MACS_kT}) and the standard deviation between the libraries (Table \ref{table: MACS_comp_2}).

\begin{table*}[h]
\centering
\caption{Experimental random (rand) and systematic (sys) relative uncertainties in this work.}
\label{table: uncertainties2}
\begin{threeparttable}
\centering
\ra{1.3}
\begin{tabular}{l c c c c c c c c c c c c c c c c c}
\toprule
 & \multicolumn{17}{c}{Uncertainty (\%)}\\
\cmidrule{2-18}
& \multicolumn{5}{c}{Exp \Rmnum{1}}&&\multicolumn{5}{c}{Exp \Rmnum{2}}&&\multicolumn{5}{c}{Exp \Rmnum{3}}\\
\cmidrule {2-6}
\cmidrule{8-12}
\cmidrule{14-18}
 & \multicolumn{2}{c}{$^{94}$Zr} && \multicolumn{2}{c}{$^{96}$Zr}& & \multicolumn{2}{c}{$^{94}$Zr} && \multicolumn{2}{c}{$^{96}$Zr}& & \multicolumn{2}{c}{$^{94}$Zr} && \multicolumn{2}{c}{$^{96}$Zr}\\
 Source of uncertainty&rand&sys&&rand&sys&&rand&sys&&rand&sys&&rand&sys&&rand&sys\\ [0.5ex]
 \midrule
target thickness, $n_t$& 0.5&&& 0.5&& & 0.4&&& 0.4&&&0.4&&&0.4\\
activated nuclei, $N_{act_{exp}}$ & 0.9&&& 0.8&&& 1.6&&& 0.6 &&&1.8&&&0.8\\
photopeak eff., $\epsilon_{\gamma}$\tnote{a} && 0.5 &&& 0.5 &&& 0.5 &&& 0.5 &&&0.5&&&0.5\\
simulation&& 1.5&&& 1.5&&& 1.5&&& 1.5&&&1.5&&&1.5\\
E$_p$, $\Delta$E$_p$ and $\Delta$z\tnote{b}&& 0.9&&& 4.5&&& 0.4&&& 2.3&&&0.6&&&1.6\\
$\sigma_{ENDF}$(Au)\tnote{c}&&1.0&&&1.0&&&1.0&&&1.0&&&1.0&&&1.0\\
$\sigma_{ENDF}$(Zr)\tnote{d}&&1.4&&&3.7&&&1.6&&&2.2&&&0.2&&&4.4\\
\midrule
Total random uncertainty&1.0 &&&0.9 &&& 1.6&&& 0.7&&&1.8&&&0.9\\
Total systematic uncertainty&&2.5 &&&6.1 &&&2.5 &&&3.7&&&2.0&&&5.0\\
\midrule
Total uncertainty &\multicolumn{2}{c}{2.7} && \multicolumn{2}{c}{6.2}& & \multicolumn{2}{c}{3.0} && \multicolumn{2}{c}{3.8}& & \multicolumn{2}{c}{2.7} && \multicolumn{2}{c}{5.1}\\ [1ex]
\bottomrule
\end{tabular}
\begin{tablenotes}
            \item[a] Uncertainty in gamma photopeak efficiency relative to the $^{198}$Au 412-keV line.
            \item[b] See separate tables \ref{table: proton energy change}, \ref{table: proton energy spread change} and \ref{table: distance change}.
            \item[c] Ref. \cite{ENDF_Unc}.
            \item[d] See equations (\ref{eq:MACS_30}) and (\ref{eq:delta_MACS_30}).
        \end{tablenotes}
     \end{threeparttable}
\end{table*}
\FloatBarrier

\section{MACS with other neutron cross section libraries}

\begin{table*}[h]
\centering
\caption{MACS (30 keV) and correction factor, $C_{lib}$, for $^{94}$Zr and $^{96}$Zr with different cross-section libraries \cite{ENDF,Jendl,Cendl,Rosfond,Jeff}.}
\label{table: MACS_comp}
\ra{1.3}
\begin{tabular}{@{}l c c c c c c c c c c c c c c c c c c@{}}
\toprule
& \multicolumn{5}{c}{Exp \Rmnum{1}}&&\multicolumn{5}{c}{Exp \Rmnum{2}}&&\multicolumn{5}{c}{Exp \Rmnum{3}}\\
\cmidrule {2-6}
\cmidrule{8-12}
\cmidrule{14-18}
& \multicolumn{2}{c}{MACS (mb)}&&\multicolumn{2}{c}{$C_{lib}$}&&\multicolumn{2}{c}{MACS (mb)}&&\multicolumn{2}{c}{$C_{lib}$}&& \multicolumn{2}{c}{MACS (mb)}&&\multicolumn{2}{c}{$C_{lib}$}\\
\cmidrule{2-3}
\cmidrule{5-6}
\cmidrule{8-9}
\cmidrule{11-12}
\cmidrule{14-15}
\cmidrule{17-18}
Library& $^{94}$Zr & $^{96}$Zr&& $^{94}$Zr & $^{96}$Zr && $^{94}$Zr & $^{96}$Zr&& $^{94}$Zr & $^{96}$Zr&& $^{94}$Zr & $^{96}$Zr&& $^{94}$Zr & $^{96}$Zr\\ [0.5ex]
 \midrule
ENDF/B-VII.1& 28.7&12.6&&0.98&1.22&&27.9&12.1&&0.96&1.18&&27.3&12.4&&0.93&1.20\\
JENDL-4.0& 27.8 & 12.5&& 1.08&1.04&&27.1&12.1&&1.05&1.00&&27.3&12.8&&1.05&1.06\\
CENDL-3.1& 28.9&12.4&&1.00&1.00&&27.8&11.8&&0.96&0.96&&27.3&12.1&&0.94&0.98\\
ROSFOND 2010&28.4&13.2&&1.00&1.20&&27.6&12.6&&0.97&1.15&&27.3&12.5&&0.96&1.14\\
JEFF-3.2&29.1&12.4&&0.99&1.00&&28.0&11.9&&0.96&0.96&&26.7&12.4&&0.91&1.00&\\
\bottomrule
\end{tabular}
\end{table*}
\FloatBarrier

\begin{table*}[h]
\centering
\caption{MACS (30 keV) of the $^{94,96}$Zr$(n,\gamma)$ reactions averaged over the three experiments and the extrapolated values at $kT$ = 8, 23 and 90 keV, using different libraries \cite{ENDF,Jendl,Cendl,Rosfond,Jeff}.}
\label{table: MACS_comp_2}
\begin{threeparttable}
\ra{1.3}
\begin{tabular}{l c c c c c c c c c c c c c}
\toprule
& \multicolumn{10}{c}{MACS (mb)}\\
\cmidrule{2-12}
$kT$ & \multicolumn{2}{c}{8 keV}&&\multicolumn{2}{c}{23 keV}&&\multicolumn{2}{c}{30 keV}&&\multicolumn{2}{c}{90 keV}\\
\cmidrule{2-3}
\cmidrule{5-6}
\cmidrule{8-9}
\cmidrule{11-12}
& $^{94}$Zr & $^{96}$Zr&& $^{94}$Zr & $^{96}$Zr && $^{94}$Zr & $^{96}$Zr&& $^{94}$Zr & $^{96}$Zr\\ [0.5ex]
 \midrule
ENDF/B-VII.1\tnote{a}& $64.1\pm3.7$&$49.8\pm5.5$&&$32.6\pm1.1$&$15.7\pm1.1$&&$28.0\pm0.6$&$12.4\pm0.5$&&$20.0\pm1.1$&$9.0\pm0.7$\\
JENDL-4.0& 67.6 & 42.5&& 33.1&15.8&&27.5&12.5&&15.4&8.0\\
CENDL-3.1& 61.5&42.8&&32.2&15.3&&28.1&12.1&&18.9&7.1\\
ROSFOND 2010&65.4&48.3&&32.9&15.6&&27.8&12.7&&17.5&11.3\\
JEFF-3.2&60.7&43.1&&31.9&15.4&&27.9&12.2&&20.5&7.7\\
\midrule
average&63.9&45.3&&32.5&15.6&&27.9&12.4&&18.5&8.6\\
st. dev.&2.8&3.5&&0.5&0.2&&0.2&0.3&&2.1&1.7\\
\midrule
this work\tnote{b}& $63.9\pm3.7$&$45.3\pm5.5$&&$32.5\pm1.1$&$15.6\pm1.1$&&$28.0\pm0.6$&$12.4\pm0.5$&&$18.5\pm2.1$&$8.6\pm1.7$\\
\bottomrule
\end{tabular}
\begin{tablenotes}
            \item[a] Uncertainties calculated using Eq. (\ref{eq:delta_MACS_30}) and (\ref{eq:delta_MACS_kT}).
            \item[b] Final uncertainties are adopted as the larger values between Eq. (\ref{eq:delta_MACS_kT}) and the standard deviation between the various libraries.
        \end{tablenotes}
     \end{threeparttable}
\end{table*}
\FloatBarrier

\clearpage
\section{Gamma spectrum from $^7$Li$(p,\gamma)^8$Be prompt capture}
\begin{figure}[h]
\centering
 \includegraphics[width=0.85\columnwidth]{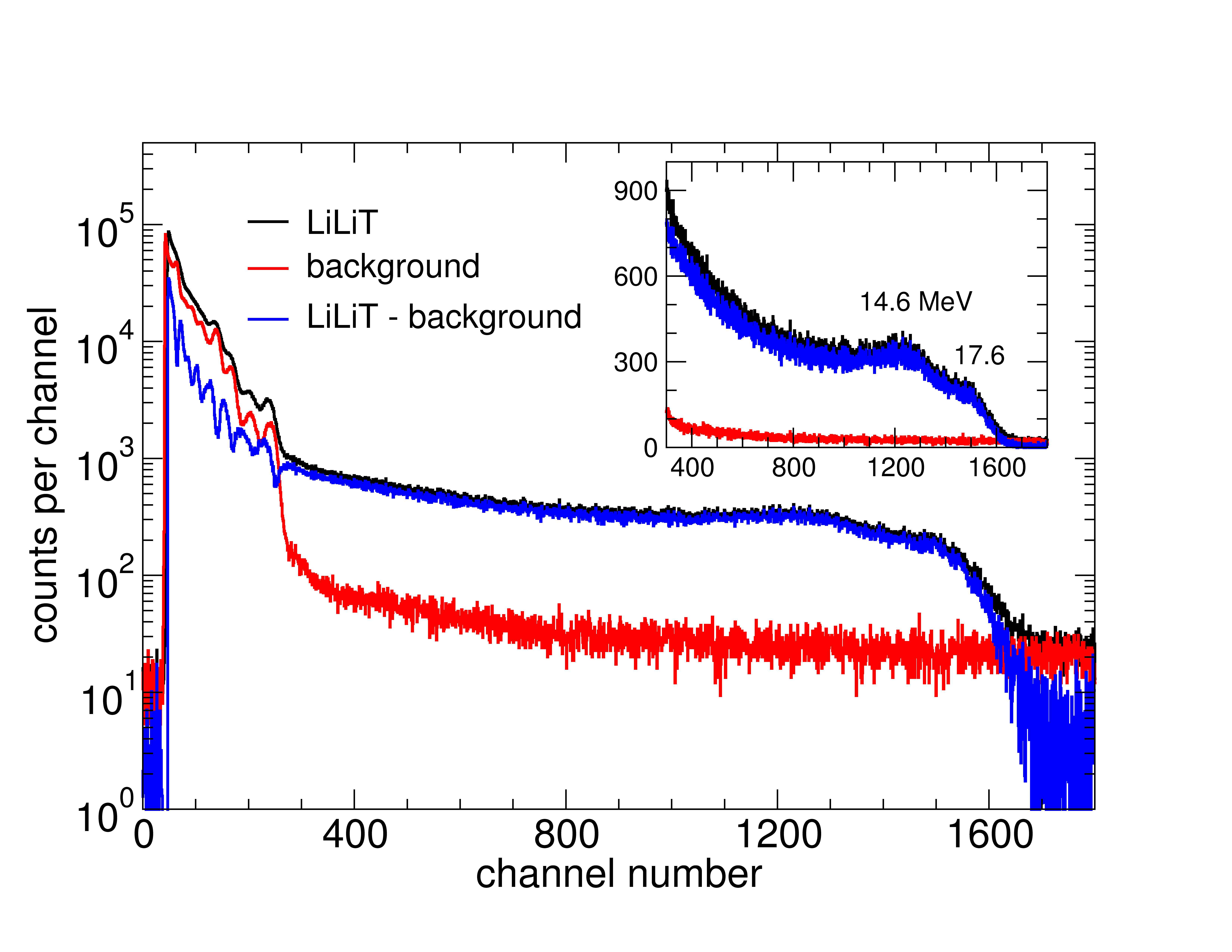}
 \caption{\label{fig: high_energy_gamma} Gamma ray spectrum from the thick-target $^7$Li$(p,\gamma)^8$Be capture reaction measured with a $\sim1.5$ mA proton beam at 1.79 MeV
 (below neutron threshold) incident on the LiLiT target. The spectrum was measured with a $6^{\prime \prime} \times 4^{\prime \prime}$ NaI(Tl) detector
 placed at a distance of 2.7 m from LiLiT behind a 1.5 m thick concrete wall.
 The black spectrum is the spectrum obtained while the proton beam impinged on the LiLiT, the red spectrum was obtained while there was no protons on LiLiT
 and the blue spectrum is the net counts.}
 \end{figure}
\FloatBarrier

\end{document}